\pdfoutput=1
\documentclass{article}
\usepackage{spconf,amsmath,graphicx}
\usepackage{amsmath}
\usepackage{amssymb}
\usepackage{bbm}
\usepackage{physics}
\usepackage{siunitx}
\usepackage{trfsigns}
\usepackage{pifont}
\usepackage{etoolbox}
\usepackage{multirow}
\usepackage{csquotes}
\usepackage[shortlabels]{enumitem}  % Allows \begin{enumerate}[label=\Alph*] ... \end{enumerate}
\usepackage[nosort]{cite}   % Groups citations like [1-4] instead of [1, 2, 3, 4]
\usepackage{hyperref}
\usepackage[capitalize]{cleveref}
\usepackage{balance} % For the balanced reference page
\usepackage{booktabs}
\usepackage[normalem]{ulem} % For underlining in aligned siunitx tables (https://tex.stackexchange.com/questions/159623/use-of-underline-and-siunitx-in-a-table)
\robustify\uline

\usepackage{url}

\newcolumntype{H}{>{\setbox0=\hbox\bgroup}c<{\egroup}@{}}   % Hidden column
\newcommand{\mrow}[1]{\multirow{2}[2]{*}{\begin{tabular}{@{}c@{}}#1\end{tabular}}}  % Centered row if using \cmidrule
\newcommand{\tab}[2][c]{\begin{tabular}{@{}#1@{}}#2\end{tabular}}   % Simple tabluar that does not add vertical spacing
\newcommand*{\thl}{\fontseries{b}\selectfont}
\robustify\thl
\robustify\fontseries
\newcommand{\setEquationSpacings}{
\abovedisplayskip4.5pt plus 3.0pt minus 1.5pt
\belowdisplayskip4.5pt plus 3.0pt minus 1.5pt
}
\makeatletter
\renewcommand\section{\@startsection {section}{1}{\z@}%
                                  {-2.75ex \@plus -1ex \@minus -.2ex}%
                                  {2ex \@plus.2ex}%
                                  {\normalfont\Large\bfseries}}
\renewcommand\subsection{\@startsection{subsection}{2}{\z@}%
                                     {-2.5ex\@plus -1ex \@minus -.2ex}%
                                     {1.2ex \@plus .2ex}%
                                     {\normalfont\large\bfseries}}
\renewcommand\subsubsection{\@startsection{subsubsection}{3}{\z@}%
                                     {-1.5ex\@plus -1ex \@minus -.2ex}%
                                     {1.0ex \@plus .2ex}%
                                     {\normalfont\normalsize\bfseries}}
\makeatother

\usepackage{tikz}
\usepackage{pgfplots}
\pgfplotsset{compat=1.16}
\usepgfplotslibrary{groupplots}
\usetikzlibrary{positioning,arrows,matrix,fit,calc,patterns,chains,scopes,shapes.multipart,decorations,decorations.markings,backgrounds}

\definecolor{palette-1}{HTML}{001C7F}    
\definecolor{palette-2}{HTML}{B1400D}    
\definecolor{palette-3}{HTML}{12711C}    
\definecolor{palette-4}{HTML}{8C0800}    
\definecolor{palette-5}{HTML}{591E71}    
\definecolor{palette-6}{HTML}{592F0D}    
\definecolor{palette-7}{HTML}{A23582}    
\definecolor{palette-8}{HTML}{3C3C3C}    
\definecolor{palette-9}{HTML}{B8850A}    
\definecolor{palette-10}{HTML}{006374}      

\tikzset{
    line/.style={draw,black,thick,rounded corners=1mm,line cap=round},
    noshortarrow/.style={line,->},
    arrow/.style={noshortarrow,shorten >=.3mm},
    doublearrow/.style={arrow,<->, shorten <=.3mm},
    box/.style={draw,black,thick,minimum height=3em,text height=1.5ex,text depth=0.25ex,rounded corners=3,fill=white},
    nopadding/.style={minimum height=0,inner sep=1mm},
    signalbox/.style={draw,black,thin,rounded corners=1mm,minimum width=7mm, minimum height=4mm,inner sep=0},
    pbox/.style={box,fill=black!10},
    backgroundbox/.style={inner xsep=3mm, inner ysep=1mm, draw, dashed, rounded corners,fill=orange!10},
    branch/.style={inner sep=0.3mm,circle,fill=black},
    operator/.style={draw,circle,black,rounded corners,inner sep=0,fill=white},
    vertex/.style={draw,ultra thin,circle,black,rounded corners,inner sep=0.6mm,fill=gray,fill opacity=0.5},
    edge/.style={line,very thick,line cap=butt},
    pattern1/.style={pattern=north west lines,pattern color=palette-1},
    pattern2/.style={pattern=north east lines,pattern color=palette-2},
    pattern3/.style={pattern=crosshatch,pattern color=palette-3},
    buswidth/.style={path picture={\draw[black,-] (path picture bounding box.south west) -- (path picture bounding box.north east);}}
}
\newcommand\defm[2]{\expandafter\newcommand{#1}{\ensuremath{#2}}} % variant without color (final)

\newcommand{\vect}[1]{\ensuremath{\boldsymbol{\mathbf{#1}}}}
\newcommand*{\tran}{^{\mkern-1.5mu\mathsf{T}}}

\newcommand{\sqnorm}[1]{\norm{#1}^2}
\newcommand{\loss}[1]{\Loss^{\text{(#1)}}}

\defm\Loss{\mathcal{L}}    % Loss
\defm\source{\vect{s}}
\defm\estimate{\hat{\source}}
\defm\noise{\vect{n}}
\defm\mix{\vect{y}}
\defm\nspk{K}
\defm\ispk{k}
\defm\tsdrthr{\tau}
\defm\tsdreps{\varepsilon}
\defm\sksdralpha{\nu}
\defm\cossim{c}
\defm\sisdralpha{\alpha} 
\defm\sisdrbeta{\beta} 
\defm\sesdreps{\varepsilon}
\defm\sdrmax{\text{SDR}_\text{max}}
\defm\reals{\mathbb{R}}

\usepackage[acronym,shortcuts]{glossaries}
\glsdisablehyper
\newacronym{SDR}{SDR}{Signal-to-Distortion Ratio}
\newacronym{CSS}{CSS}{Continuous Speech Separation}
\newacronym{PIT}{PIT}{Permutation Invariant Training}
\newacronym{uPIT}{uPIT}{Utterance-level \gls{PIT}}
\newacronym{MSE}{MSE}{Mean Squared Error}
\newacronym{DFS}{DFS}{Depth First Search}
\newacronym{DPRNN}{DPRNN}{Dual-Path Recurrent Neural Network}
\newacronym{WER}{WER}{Word Error Rate}
\newacronym{SSE}{SSE}{Sum Squared Error}
\newacronym{SI-SDR}{SI-SDR}{Scale-Invariant \gls{SDR}}
\newacronym{DP}{DP}{Dynamic Programming}
\newacronym{tSDR}{tSDR}{thresholded \gls{SDR}}
\newacronym{aSDR}{A-SDR}{averaged \gls{SDR}}
\newacronym{saSDR}{SA-SDR}{source-aggregated \gls{SDR}}
\newacronym{DER}{VAER}{Voice Activity Error Rate}
\newacronym{TasNet}{TasNet}{Time-domain Audio Separation Network}

\title{SA-SDR: A novel loss function for separation of meeting style data}
\newcommand{\upb}{$^1$}
\newcommand{\ntt}{$^2$}
\name{\begin{tabular}{c}Thilo von Neumann\upb, Keisuke Kinoshita\ntt, Christoph Boeddeker\upb, Marc Delcroix\ntt, \\ Reinhold Haeb-Umbach\upb\end{tabular}}
\address{\upb Paderborn University, Germany \quad \ntt NTT Corporation, Japan}
\begin{document}
\ninept
\setEquationSpacings
\maketitle
\begin{abstract}
% 100-150 words
Many state-of-the-art neural network-based source separation systems use the averaged Signal-to-Distortion Ratio (SDR) as a training objective function. The basic SDR is, however, undefined if the network reconstructs the reference signal perfectly or if the reference signal contains silence, e.g., when a two-output separator processes a single-speaker recording. Many modifications to the plain SDR have been proposed that trade-off between making the loss more robust and distorting its value. We propose to switch from a mean over the SDRs of each individual output channel to a global SDR over all output channels at the same time, which we call source-aggregated SDR (SA-SDR). This makes the loss robust against silence and perfect reconstruction as long as at least one reference signal is not silent. We experimentally show that our proposed SA-SDR is more stable and preferable over other well-known modifications when processing meeting-style data that typically contains many silent or single-speaker regions.
\end{abstract}
\section{Introduction}
\label{sec:intro}

%% Separation is important
Source separation is an important pre-processing step for many other systems, such as speech recognition or diarization, that often cannot handle recordings of overlapping speech.
Advances in the past years using neural network-based source separators have led to impressive results on fully overlapped clean anechoic recordings \cite{Luo2018_TaSNetTimeDomainAudio,Luo2020_DualPathRNNEfficient,Kolbaek2017_MultitalkerSpeechSeparationa,Subakan2021_AttentionAllYou}.
More realistic and challenging scenarios like meeting-style data recently gained research interest \cite{Chen2020_ContinuousSpeechSeparation,Chen2021_ContinuousSpeechSeparation,Neumann2019_AllneuralOnlineSource,Carletta2006_AMIMeetingCorpus,Anguera2012_SpeakerDiarizationReview,VanSegbroeck2020_DiPCoDinnerParty,Watanabe2020_CHiME6ChallengeTackling}, where speakers do not fully overlap and a separation system has to handle a varying number of speakers including silence.

% Aus RSAN-Paper noch ein paar Zitate holen? Klingt gut
% Man könnte noch AMI, Chime... zitieren
% Auch gut. Als motivation. Allerdings haben die hall.
% DiPCo--Dinner Party Corpus  - ALternativer CHiME-5 eval datesatz
% CHiME 5 oder 6? Oder sehe ich die gerade nicht

%% SDR loss and its problems
Many state-of-the-art separation systems, like the \gls{TasNet} \cite{Luo2018_TaSNetTimeDomainAudio,Luo2020_DualPathRNNEfficient}, maximize the \gls{SDR} as the objective during training.
However, the standard \gls{SDR} becomes problematic (1) when the system is asked to reconstruct silence, as it is often required in realistic meeting-style conversations when one speaker listens while another utters, and (2) when it reconstructs one reference signal very well.
In both cases, the value of the \gls{SDR} explodes.

%% Solutions from the literature
Many works address this problem by modifying the \gls{SDR} for each estimated separated signal \cite{Wisdom2020_UnsupervisedSpeechSeparation,vonNeumann2021_GraphPITGeneralizedPermutation,vonNeumann2020_MultiTalkerASRUnknown,Wisdom2021_WhatAllFuss,Luo2020_SeparatingVaryingNumbers,Heitkaemper2020_DemystifyingTasNetDissecting}.
To address the instabilities for perfect reconstruction, one can limit the value range of the \gls{SDR} by introducing a soft maximum \cite{Wisdom2020_UnsupervisedSpeechSeparation} or by skewing its curve \cite{Luo2020_SeparatingVaryingNumbers}.
Instabilities due to silent targets can be addressed by switching to a log-\gls{MSE} variant \cite{Heitkaemper2020_DemystifyingTasNetDissecting,Wisdom2021_WhatAllFuss} or by adding small values to the fraction in the \gls{SDR} \cite{vonNeumann2021_GraphPITGeneralizedPermutation}.
All of these modifications distort the loss value for each separated speech signal.

%% SA-SDR: Modify the aggregation
We propose not to modify the \gls{SDR} definition for each output channel, but the way it is aggregated across outputs.
The common way of aggregation is a simple arithmetic mean over the individual \glspl{SDR} of each output, --- the \gls{aSDR}, e.g., \cite{Kolbaek2017_MultitalkerSpeechSeparationa,Luo2018_TaSNetTimeDomainAudio,Subakan2021_AttentionAllYou}.
We propose to transition from these \enquote{local} \glspl{SDR} to a \enquote{global} \gls{SDR} that combines all outputs to one long signal before computing the \gls{SDR}.
This is done by summing the energies of all targets and all error terms --- the \gls{saSDR}.

%% Results
We found experimentally that the proposed \gls{saSDR} achieves one of the best performances among the presented losses, measured with various metrics, and comes without any hyperparameters to tune.
Making the loss robust against silence is important for training on realistic meeting-style data where such a case  frequently occurs so that more training data can be used.
Limiting the value range of the \gls{SDR} in general improves the performance of the trained models.
% This allows us to use more training data which improves the overall performance.
We additionally propose to use \gls{saSDR} as a signal-level evaluation metric for meting-style data where the classical \gls{SDR} cannot be computed.
The \gls{saSDR} measures in one metric both how well active and silent sources are estimated.

% \begin{itemize}
%     \item Source separation is important!
%     \item State-of-the-art uses SDR-based losses
%     \item Issues: becomes unstable for perfect reconstruction or silence target
%     \item Literature Solutions: threshold, skewing, eps, ...
%     \item Proposed: Modify the way the SDR is aggregated among output channels
%     \item Results: No performance degradation on fully overlapped data; better suppression in silent regions of meeting style data
% \end{itemize}

\glsresetall % I want the A-SDR and SA-SDR to be defined again on next usage

\section{Conventional loss functions: SDR and its variants}
\label{sec:conventional}

We consider speech mixtures $\mix\in\reals^T$ of $\nspk$ speakers.
A mixture signal $\mix = \sum_{\ispk = 1}^\nspk \source_\ispk + \noise$ is the sum of the speech of individual speakers $\source_\ispk\in\reals^T$ and noise $\noise\in\reals^T$.
All signals are represented as vectors of samples with a time length of $T$.

The process of obtaining estimates $\estimate_\ispk$ for the clean reference signals $\source_\ispk$ from the mixture $\mix$ is called source separation.
The estimates $\estimate_\ispk$ should reconstruct the clean signals $\source_\ispk$ as closely as possible up to a permutation between estimates and references.
% It is thus trained with a \gls{PIT} scheme \cite{Kolbaek2017_MultitalkerSpeechSeparationa} using a time-domain objective function.

% Source separation is a process of extracting each speaker's individual speech from a mixture of multiple speakers.
% A mixture signal $\mix = \sum_{\ispk = 1}^\nspk \source_\ispk + \noise$ is the sum of the speech of individual speakers $\source_\ispk$ and noise $\noise$.

The \gls{SDR} -- and variations of it -- is a commonly used training objective and evaluation metric for such source separation models.
In its basic form, it is defined for a pair of an estimated signal $\estimate$ and a corresponding reference signal $\source$:
\begin{align}
    \text{SDR}(\estimate,\source) = 10 \log_{10} \frac{\sqnorm{\source}}{\sqnorm{\estimate - \source}}.
\end{align}
The estimation error $\estimate - \source$ is to be minimized at the output of a source separator, so the objective to be minimized becomes the negative \gls{SDR}, $\loss{SDR}=-\text{SDR}$, for each output channel.

We only consider scale-dependent \glspl{SDR} here, but the same conclusions could be made with scale-invariant losses by re-scaling the target as $\source^\text{(re-scaled)} = \source \frac{\source\tran\estimate}{\sqnorm{\source}}$.
% The re-scaling makes the gradients perpendicular to $\estimate$.

The plain \gls{SDR} is undefined if the target signal is silent ($\source = \vect{0}$), when, e.g., a two-output separator is trained to process a single-speaker utterance, or if the reconstruction is perfect ($\estimate = \source$).
Even if these edge-cases are not hit, the \gls{SDR} explodes if the reference is close to $\vect 0$ or the estimation is almost perfect.
It is often desired to train with silent references, especially with realistic training data, and perfect reconstruction should never be a problem.

The remainder of this section discusses different modifications to the plain \gls{SDR} that make it robust.
Just preventing the loss value from exploding for silent references often just moves the problem to a later point in training.
Since a network can trivially estimate silence, it can easily learn to reconstruct silence (almost) perfectly, so the loss additionally needs to counter perfect reconstruction.

\subsection{Soft Maximum}

One way to make the \gls{SDR} robust against perfect reconstruction is to impose a soft maximum with the \gls{tSDR} \cite{Wisdom2020_UnsupervisedSpeechSeparation}:
\begin{align}
    \loss{tSDR} = -10 \log_{10} \frac{\norm{\source}^2}{\norm{\source - \estimate}^2 + \tsdrthr \norm{\source}^2},
    \label{eq:tsdr}
\end{align}
where $\tsdrthr = 10^{-\sdrmax / 10}$.
It can be made robust against silence by adding a small constant $\tsdreps>0$ to the reference signal \cite{vonNeumann2021_GraphPITGeneralizedPermutation}:
% \fromcb{Why smaller 1? I don't see a reason for an upper bound. TvN: I wanted to show that eps is small}
%
\begin{align}
    \loss{\tsdreps-tSDR} = -10 \log_{10} \frac{\norm{\source}^2 + \tsdreps}{\norm{\source - \estimate}^2 + \tsdrthr (\norm{\source}^2 + \tsdreps)}.
\end{align}
%
% The $\tsdreps$ does not change the value of $\loss{\tsdreps-tSDR}$ for perfect reconstruction.
% The \tsdreps-tSDR is robust against silent targets and still interpretable.
% The gradient points into the same direction as the gradient of the plain \gls{SDR} but is scaled differently:
% %
% \begin{align}
%     \nabla_\estimate\text{\tsdreps-tSDR} = \frac{20}{\ln10} \frac{\estimate - \source}{\norm{\estimate - \source}^2 + \tsdrthr(\sqnorm{\source} + \tsdreps)}
% \end{align}
% %
Note that both $\tsdrthr$ and $\tsdreps$ do not influence the direction of the gradient of $\loss{\tsdreps-tSDR}$ but only its scaling and the ratio between different output channels.
The $\tsdreps$-tSDR thus gives a smaller weight to the output channels and examples that are well separated.

\subsection{Skewing the SDR}

Another variation of \gls{SDR} that tries to combat the numerical instabilities for perfect reconstruction is the skewed SDR \cite{Luo2020_SeparatingVaryingNumbers}.
% It is originally formulated in a scale-invariant way with the cosine similarity $\cossim$:
% %
% \begin{align}
%     \loss{skewed SI-SDR} = -10\log_{10}\frac{\cossim(\source,\estimate)}{1-\cossim(\source,\estimate) + \sksdralpha}.
% \end{align}
% %
% Note that $\sksdralpha$ does not impose a general soft maximum on the loss value.
% We can remove the scale-independence by formulating it the same way as the thresholded SDR:
It is originally formulated in a scale-invariant way, but we only consider the scale-dependent variant:
\begin{align}
    \loss{skewed SDR} = -10 \log_{10} \frac{\norm{\source}^2}{\norm{\source - \estimate}^2 + \sksdralpha\norm{\estimate}^2}
\end{align}
The skewing factor $\sksdralpha > 0$ controls how much the loss value is skewed for small reconstruction errors.
The additional term $\sksdralpha\sqnorm{\estimate}$ pushes the estimation towards $\vect 0$ in the scale-dependent variant.\footnote{This effect is not present in the scale-invariant variant, but we only consider scale-dependent losses here.}
It is unclear if it is well-suited for source separation.
% This behavior can be observed from the gradient:
% Gradient:
% %
% \begin{align}
%     \nabla_\estimate\mathrm{skewed SDR} = \frac{20}{\ln10} \frac{(1 + \sksdralpha)\estimate - \source}{\norm{\estimate - \source}^2 + \sksdralpha\norm{\estimate}^2}
% \end{align}
%
% The direction of the gradient is bend towards $0$ by $\sksdralpha$ \inred{?}.

% This loss is only used for auto-encoding in \cite{Luo2020_SeparatingVaryingNumbers} but not for separation; it is unclear if this modification is well suited as an objective function for source separation.

\subsection{log-MSE}
\label{sec:log-mse}

A simple way to avoid the instability for silent targets is to ignore the numerator $\sqnorm{\source}$.
This leads to the log-MSE loss \cite{Heitkaemper2020_DemystifyingTasNetDissecting}:
\begin{align}
    \loss{log-MSE} = \log_{10} \sqnorm{\source - \estimate}.
\end{align}
It has the same gradients as $\loss{SDR}$ but scaled differently.
As discussed earlier, the loss additionally has to be made robust against perfect reconstruction, e.g., by adding a constant to the argument of the logarithm \cite{vonNeumann2020_MultiTalkerASRUnknown}:
\begin{align}
    \loss{log1p-MSE} = \log_{10} (\sqnorm{\source - \estimate} + 1).
\end{align}
The log-MSE loss has the disadvantage that its value depends on the scaling of the signals and thus varies more and is more difficult to interpret than the \gls{SDR}.
Especially when the best model is selected based on the development loss, a sub-optimal model might be selected.
% \inred{CB: Conclusion: Follow the sugguestions of any ML book and calculate a metric to do the model selection? Regarding the scale: It does not matter, because it is a constant and can be consumed from the learning rate. TVN: I'd just ignore this issue here}
% 
Similar modifications are possible as for the SDR-based variants, such as adding a soft minimum similar to \cref{eq:tsdr} \cite{Wisdom2021_WhatAllFuss}.

\subsection{Extra Loss for Silence}

A different way to handle problematic inputs is to identify them and use an alternative loss where the \gls{SDR} is not applicable.
One example for this is using the mixture signal $\mix$ instead of the target in a thresholded loss where the target is silent but the mixture is not \cite{Wisdom2021_WhatAllFuss}, here as a variant of the log-tMSE:
\begin{align}
    \Loss_0^\text{(log-tMSE)} = 10\log_{10} ( \sqnorm{\estimate} + \tsdrthr\sqnorm{\mix}).
\end{align}
This loss is only applied where the target is silent, i.e., $\source=\vect 0$.
Applying different losses to different outputs can create discontinuities in the gradients.
Besides that, the decision which outputs are silent is not always trivial, e.g., for very short segments of speech.

% \subsubsection{SE-SI-SDR}
% \inred{I don't have a model for this loss!}
% Proposed in a paper at interspeech

% \begin{align}
%     \loss{SE-SI-SDR} = -10\log_{10} \frac{\sqnorm{\alpha\source} + \sesdreps}{\sqnorm{\alpha\source - \estimate} + \sesdreps}
% \end{align}

% The scale-dependent variant:

% \begin{align}
%     \loss{SE-SDR} = -10\log_{10} \frac{\sqnorm{\source} + \sesdreps}{\sqnorm{\source - \estimate} + \sesdreps}
% \end{align}

% Possible problem: Perfect value for silent target is 0 while for non-silent targets the optimal value is $-10\log_{10}\sesdreps \rightarrow -\infty$.

% \subsubsection{Generalized SDR loss}

% \inred{Is this needed?}We can combine the SDR, tSDR and skewed SDR losses into one:

% \begin{align}
%     \text{SDR} = -10\log_{10}\frac{\sqnorm{\source} + \tsdreps}{\sqnorm{\source - \sisdrbeta\estimate} + \tsdrthr(\sqnorm{\source} + \tsdreps) + \sksdralpha\sqnorm{\sisdrbeta\estimate}}
% \end{align}

\section{Aggregating SDR across outputs}
The modifications discussed so far all modify the \gls{SDR} for each individual output.
But, a source separator has multiple outputs and the losses for different outputs have to be combined.
Most source separation techniques that use an \gls{SDR}-based loss average the loss over the output channels, e.g., \cite{Luo2018_TaSNetTimeDomainAudio,Luo2020_DualPathRNNEfficient,Chen2020_ContinuousSpeechSeparation,Wisdom2020_UnsupervisedSpeechSeparation,vonNeumann2021_GraphPITGeneralizedPermutation,Heitkaemper2020_DemystifyingTasNetDissecting}.
For the standard \gls{SDR}, this can be written as
\begin{align}
    \loss{A-SDR} 
    = \frac{1}{10}\sum_{k=1}^K \Loss(\source_\ispk, \estimate_\ispk)
    = -\frac{10}{\nspk}\sum_{\ispk = 1}^{\nspk} \log_{10} \frac{\sqnorm{\source_\ispk}}{\sqnorm{\source_\ispk - \estimate_\ispk}}.
\end{align}
Extensions to all other single-channel losses described in \cref{sec:conventional} are straightforward.
We call this conventional way of combining the single-channel \glspl{SDR} the \gls{aSDR}. 
It suffers from the aforementioned problems with the standard \gls{SDR}:
It becomes unstable if any output channel has perfect reconstruction or a silent reference signal.
% \fromcb{Note sure, maybe add something like: Especially the silent target signals is likely, because it means, that one speaker utters, while the other listens. TvN: Added that to the introduction}

We propose to stabilize the loss by, instead of computing the arithmetic mean, summing the energies of the targets and distortions:
\begin{align}
\loss{SA-SDR} = -10\log_{10}\frac{\sum_{\ispk=1}^\nspk \norm{\source_\ispk}^2}{\sum_{\ispk=1}^\nspk \norm{\source_\ispk - \estimate_\ispk}^2}.
\end{align}
This is equivalent to concatenating all output channels to compute a global \gls{SDR} and we call it \gls{saSDR}.
It is stable as long as at least one reference is not perfectly reconstructed and at least one is not completely silent.
The case of complete silence, i.e., all reference signals are zero, is not considered here since separation is trivial in that case and silence can easily be detected.

Both \gls{aSDR} and \gls{saSDR} are aggregations over the \glspl{SDR} of the individual output channels and thus bounded by them, i.e.,
\begin{align}
    \min_k \text{SDR}(\estimate_k, \source_k) \leq \text{\phantom{S}A-SDR} &\leq \max_k \text{SDR}(\estimate_k, \source_k), \\
    \min_k \text{SDR}(\estimate_k, \source_k) \leq \text{SA-SDR} &\leq \max_k \text{SDR}(\estimate_k, \source_k).
\end{align} 
From this follows that for a special case where the \glspl{SDR} of all individual output channels are equal ($\text{SDR}(\estimate_1, \source_1)=\text{SDR}(\estimate_2, \source_2)=...$), \gls{aSDR} and \gls{saSDR} are also equal.

\subsection{Energy of the Reference Signals}
The \gls{aSDR} weights each output channel equally, independent of its energy level.
This is often not desired: When a reference signal contains only a short segment of speech (i.e., low energy), it gets weighted the same as a longer speech signal in another output channel.
This gives the samples in the short speech fragment an extraordinarily large weight.
The \gls{saSDR} is less sensitive to these outliers as it implicitly weights the output channels by their energy and focuses less on low-energy signals.

\subsection{Energy of the Distortions}
Having a single well-separated output signal $\estimate_l$ is enough to push the \gls{aSDR} to extremely good values even if other outputs are separated poorly.
The \gls{aSDR} thus focuses the already well separated outputs while the \gls{saSDR} minimizes the total distortions by focusing the poorly separated outputs.
This can be seen from the gradients.

The gradient of the $l$-th output $\estimate_l$ of \gls{aSDR} depends only on the $l$-th output signal
\begin{align}
    \nabla_{\estimate_l} \loss{A-SDR} = \frac{20}{\nspk\ln 10} \frac{\estimate_l - \source_l}{\sqnorm{\estimate_l - \source_l}},
\end{align}
while the gradients of \gls{saSDR} depend on all output signals:
\begin{align}
    \nabla_{\estimate_l} \loss{SA-SDR} = \frac{20}{\ln 10} \frac{\estimate_l - \source_l}{\sum_k \sqnorm{\estimate_k - \source_k}}.
\end{align}
One would expect the gradients of the output with worse quality to be larger, i.e., $\norm{\nabla_{\estimate_k}  \loss{A-SDR}} > \norm{\nabla_{\estimate_l}  \loss{A-SDR}}$ if $\sqnorm{\estimate_k - \source_k} > \sqnorm{\estimate_l - \source_l}$.
But the opposite is true for \gls{aSDR}:
\begin{align}
    \frac{\norm{\nabla_{\estimate_k} \loss{A-SDR}}}{\norm{\nabla_{\estimate_l} \loss{A-SDR}}} = \frac{\norm{\estimate_l - \source_l}}{\norm{\estimate_k - \source_k}} < 1 \text{ if $l$ is better separated}.
\end{align}
The \gls{saSDR} has the expected behavior:
\begin{align}
    \frac{\norm{\nabla_{\estimate_k} \loss{SA-SDR}}}{\norm{\nabla_{\estimate_l} \loss{SA-SDR}}} = \frac{\norm{\estimate_k - \source_k}}{\norm{\estimate_l - \source_l}} >1 \text{ if $l$ is better separated}.
\end{align}
%
% \gls{aSDR} focuses the already well separated outputs while \gls{saSDR} gives a larger weight to the worse channel.
% This means that \gls{aSDR} can be pushed to extreme values by only separating a single speaker well and ignoring all other speakers.
% \inred{It is unclear how strong this effect becomes with sophisticated optimizers, like Adam, that normalize the gradients.}

The \gls{saSDR} is not only an elegant way to make the \gls{SDR} robust against silent targets and perfect reconstruction in common use-cases where some speakers make a pause, it also leads to a better balance between the output channels.

\section{Experiments}

% We conduct experiments to show the effect of the mentioned modifications to the \gls{SDR} loss in this section. \inred{RH: sentence can be removed to save space.}

\subsection{Data}
We evaluate the different loss functions on fully overlapped mixtures from the WSJ0-2mix database \cite{Hershey2016_DeepClusteringDiscriminative} and on artificially generated meetings \cite{vonNeumann2021_GraphPITGeneralizedPermutation} based on WSJ \cite{Garofalo2007_CsriWsj0Complete}.
Each meeting is about \SI{120}{\second} long, contains 5-8 speakers, an overlap ratio between $0.2$ and $0.4$ and is corrupted by white microphone noise of \SIrange{20}{30}{\decibel}.
Following the ideas of \gls{CSS} \cite{Chen2020_ContinuousSpeechSeparation}, there are never more than two speakers overlapping at the same time.

We randomly cut \SI{2}{\second} long segments from the meetings for training.
This segment size was shown to work well on this data in \cite{vonNeumann2021_GraphPITGeneralizedPermutation}.

\subsection{Model Training}
%% Network and training
We use a \gls{DPRNN} \cite{Luo2020_DualPathRNNEfficient} with two outputs and the default configuration from \cite{Luo2020_DualPathRNNEfficient} for experiments on fully overlapped data, i.e., six blocks, a feature size of $64$ and a window size and shift of $100$ and $50$, respectively.
To speed up our experiments on meeting-style data, we use a shallower model with only three blocks.
We train all models with \gls{PIT} for the same number of iterations with the same batch size.
We pick the best checkpoint for evaluation based on the loss on the development set.

%% Meeting-style data
% We exclude any single-speaker segments for the losses that cannot handle them (see column \enquote{\#spk train} in \cref{tab:meetings}).

%% Stitching
We use a stitching approach \cite{Chen2020_ContinuousSpeechSeparation,vonNeumann2021_GraphPITGeneralizedPermutation} to evaluate our model on the \SI{120}{s} long meetings.
The input signal is segmented into overlapping segments of \SI{2.4}{\second} length, each segment is processed by the separator, and the separated signals from adjacent segments are aligned to minimize the mean squared error between the overlapping signal parts.
The stitcher uses a future and history context of \SI{1}{\second} each.

%% Loss parameters
We choose $\sksdralpha=0.3$ for the skewed losses, and set $\sdrmax=\SI{30}{\decibel}$ and $\tsdreps=10^{-6}$ for the thresholded losses (prefixed with \enquote{t}).

\subsection{Metrics}

\subsubsection{Word Error Rate (WER)}%%\vspace{-1em}
To obtain a \gls{WER}, we use a speech recognizer from the ESPnet toolkit \cite{Watanabe2018_ESPnetEndtoEndSpeech} trained on clean WSJ data.
It achieves a \gls{WER} of \SI{5.6}{\percent} on the clean eval92 set of WSJ.

We do not compute the \gls{WER} for the full meetings because of two reasons:
The speech recognizer poorly generalizes to long signals and the alignment of estimated transcriptions with the ground truth is difficult.
We therefore cut the separated signals at the ground truth utterance boundaries and compute the average \gls{WER} over these utterances, choosing the output channel with the lower \gls{WER}.
This explicitly ignores regions in the output that should be silent.
% \fromcb{Same as in CSS paper? TvN: No, but I don't want to discuss that here}

\subsubsection{Signal-to-distortion Ratio (SDR)}%%\vspace{-1em}

As a signal-level metric, we compute the BSSEval-SDR \cite{Vincent2006_PerformanceMeasurementBlind,Raffel2014_MirEvalTransparent}.
Similar to \gls{WER}, it is not meaningful to compute the BSSEval-SDR over a whole output signal for meeting-style data
% because it assumes a constant room impulse response for each signal which is not given when different speakers appear on each output channel.
because each output channel can contain more than one utterance and processing one output channel would follow the source-aggregated idea while the channels are averaged, i.e., the aggregation would be a mixture of source-aggregation and averaging.
BSSEval-SDR usually uses averaging, hence we use the same processing as for \gls{WER}: We cut utterances from the separated signals and compute the BSSEval-SDR for each utterance independently.
For WSJ0-2mix, we compute the BSSEval-\gls{SDR} over the full signals using the \emph{min} sub-set.

\subsubsection{Attenuation Ratio for Silence}%%\vspace{-1em}
To judge how well the systems can suppress speech where the output should be silent, we compute an attenuation ratio
\begin{align}
    \text{attenuation-ratio} = 10\log_{10}\frac{\sqnorm{\mix^\text{(sil)}}}{\sqnorm{\estimate^\text{(sil)}}},
\end{align}
where $\mix^\text{(sil)}$ and $\estimate^\text{(sil)}$ are the signal parts that should be silent in the mixture and separated streams, respectively.
When the evaluated system favors a suppression, e.g. as the skewed SDR, the value may be overoptimistic for those systems.

\subsubsection{Voice Activity Error Rate (VAER)}%%\vspace{-1em}
We use webrtcvad\footnote{\url{https://github.com/wiseman/py-webrtcvad}} to obtain hypotheses for speech activity from the separated signals.
From these, we compute a \gls{DER} by comparing the estimated speech activity with the ground truth overlap-free voice activity labels using pyannote \cite{pyannote.metrics}.
This metric has the advantage compared to the \gls{WER}, \gls{SDR} and attenuation ratio that it judges the quality of the full output streams including silence.
It, however, only judges how well the system can discriminate where speakers are active and not the separation quality in general.

% \fromcb{Not sure, but maybe some details (Depends on the complexity of the tool). e.g. tolerance of the onset and offset estimates} \inred{Used the default settings for pyannote: collar = 0, don't ignore overlaps; for webrtcvad: 80 samples frame length, default mode}

\subsubsection{SA-SDR with Graph-PIT}%%\vspace{-1em}
As a signal-level metric that measures the overall quality of a system output, we propose to use the \gls{saSDR}.
We use the ideas from Graph-PIT\footnote{\url{https://github.com/fgnt/graph_pit}} \cite{vonNeumann2021_GraphPITGeneralizedPermutation,vonNeumann2020_SpeedingPermutationInvariant} to construct reference signals for the output streams for meeting-style data from the reference utterance signals because the placement of utterances on output channels is irrelevant.
\gls{aSDR} is not well applicable here for the same reasons as BSSEval-SDR while \gls{saSDR} does not depend on the placement of utterances.
The optimal assignment of utterances to output channels for the reference signal is much more efficient to compute for \gls{saSDR} than for \gls{aSDR} \cite{vonNeumann2020_SpeedingPermutationInvariant}.

\subsection{A-SDR and SA-SDR on Fully Overlapped Data}

\begin{table}[t]
    \centering
    \caption{Comparison of the separation performance of \gls{aSDR} and \gls{saSDR} on WSJ0-2mix. Separation performance is evaluated with BSS-eval SDR \cite{Raffel2014_MirEvalTransparent}.
    % \inred{Why is WER so bad for SA-SDR models? This doesn't make sense, the WER correlates pretty well with the SDR for meeting-like data}
    }
    \begin{tabular}{lSSSH}
    \toprule
    Loss & {\tab{BSSEval\\SDR}} & {A-SDR} & {SA-SDR} & {WER\footnotemark} \\
    \midrule
    no separation & 0.2 & 0.0 & 0.0 & \\
    \midrule
     A-SDR \cite{Luo2018_TaSNetTimeDomainAudio,Luo2020_DualPathRNNEfficient} & 17.8 & 17.5 & 17.8 & 14.2 \\    % upit/sunny_gecko
     A-tSDR \cite{Wisdom2020_UnsupervisedSpeechSeparation} & 17.8 & 17.5 & 17.8 & 12.1 \\   % upit/unlikely_squid
     SA-SDR & \thl 18.0 & \thl 17.7 & \thl 18.0 & 14.4\\ % upit/burning_warbler
     SA-tSDR & 17.7 & 17.5 & 17.8 & 15.2 \\  % upit/tragic_marmot
     \bottomrule
\end{tabular}

% \begin{tabular}{l SSSSS}
%     \toprule
%     \multirow{2}[1]{*}{Metric} & & & Loss & \\
%     \cmidrule(lr){2-6}
%     & {input} & {a-SDR} & {a-tSDR} & {sa-SDR} & {sa-tSDR} \\
%     \midrule
%     BSSEval SDR & 0.16 & 17.7 & 17.8 & \thl 18.0 & 17.7 \\
%     SDR & & & & & \\
%     sa-SDR & & & & & \\
%     WER & & & & & \\
%     \bottomrule
% \end{tabular}
    \label{tab:wsj0_2mix}
\end{table}
% \footnotetext{WER is computed on the \emph{max} subset while all other metrics are computed on \emph{min}.}

To show that the \gls{saSDR} is well suited for general source separation purposes, we first compare it with the conventional \gls{aSDR} on the common task to separate fully overlapped speech from the WSJ0-2mix database in \cref{tab:wsj0_2mix}.
The model trained with \gls{saSDR} performs slightly better than the model trained with \gls{aSDR}, while the variants with a threshold, A-tSDR and SA-tSDR, show a comparable performance.
The value of SA-SDR as a metric is close to BSSEval-SDR and A-SDR for this data.
It is hence an alternative metric for source separation in clean anechoic scenarios.

% \inred{Mention that these are better numbers than the original paper with SI-SNR and the same network configuration? -> No, not enough space and the original paper probably trained shorter}
% \fromcb{I found 19.0 in the DPRNN paper for SDRi.} \inred{Yes, but that is a larger model / smaller window shift. They got 16.2dB SDRi for this configuration}

\subsection{A-SDR and SA-SDR on Meeting-Style data}
\begin{table}[t]
    \centering
    \caption{Comparison of the separation performance of SDR variants on meeting-style data. Averaged losses are prefixed with \enquote{A-} and source-aggregated losses with \enquote{SA-}. Best numbers are \textbf{bold} and best numbers among conventional averaged SDRs are \uline{underlined}.}
    \def\aggs{source}
\def\agga{average}
\def\emptycell{{---}}
\def\twospk{2}
\def\onetwospk{1+2}
\addtolength{\tabcolsep}{-4pt}
\robustify\uline
\def\Uline#1{#1\llap{\uline{\phantom{#1}}}}  % https://tex.stackexchange.com/a/258867/148912
% \robustify\tnote
\begin{tabular}{lHcSSSSS}
    \toprule
    \mrow{Loss}&\mrow{aggregation} &\mrow{\tab{\#spk\\train}}& \multicolumn{5}{c}{Metrics}\\
    \cmidrule{4-8}
     &  &  & {WER} & {\tab{atten.\\ratio}} & {\tab{BSSEval\\SDR}} &{VAER} & {\tab{SA-\\SDR}} \\
    \midrule
    no separation & \emptycell & \emptycell & 48.1 & 0.0 & 7.3 & 65.6 & 0.0 \\
    \midrule
A-SDR \cite{Luo2018_TaSNetTimeDomainAudio}  & \agga & \twospk & 13.5 & 25.5 & 19.1 & 12.6 & 13.8 \\ %% upit_generalized_sdr/final_perch/ckpt_530000_SMSWsjMeetingReader_stitcher_8000_3200_8000_1
A-log-MSE \cite{Heitkaemper2020_DemystifyingTasNetDissecting} & \agga & \twospk & 13.1 & 18.3 & 19.5 & 13.2 & 14.8 \\ %% upit_generalized_sdr/surprised_kite/ckpt_880000_SMSWsjMeetingReader_stitcher_8000_3200_8000_1
A-log1p-MSE \cite{vonNeumann2020_MultiTalkerASRUnknown} & \agga & \onetwospk & 13.5 & 25.3 & \Uline{19.6} & \Uline{9.9} & \Uline{16.8} \\ %% upit_generalized_sdr/visiting_carp/ckpt_620000_SMSWsjMeetingReader_stitcher_8000_3200_8000_1
A-skewed-SDR \cite{Luo2020_SeparatingVaryingNumbers} & \agga & \twospk & 15.6 & 24.7 & 18.7 & 12.5 & 10.1 \\ %% upit_generalized_sdr/busy_guineafowl/ckpt_550000_SMSWsjMeetingReader_stitcher_8000_3200_8000_1
% A-skewed-tSDR  & \agga & \twospk & 15.0 & 22.7 & 18.8 & 11.9 & 10.6 \\ %% upit_generalized_sdr/unnecessary_bear/ckpt_410000_SMSWsjMeetingReader_stitcher_8000_3200_8000_1
A-tSDR \cite{Wisdom2020_UnsupervisedSpeechSeparation} & \agga & \twospk & 13.6 & 21.1 & 18.8 & 14.0 & 13.3 \\ %% upit_generalized_sdr/pleased_caribou/ckpt_690000_SMSWsjMeetingReader_stitcher_8000_3200_8000_1
A-\tsdreps-tSDR \cite{vonNeumann2021_GraphPITGeneralizedPermutation} & \agga & \onetwospk & \Uline{12.8} & 25.9 & \Uline{19.6} & 11.8 & 15.5 \\ %% upit_generalized_sdr/massive_duck/ckpt_890000_SMSWsjMeetingReader_stitcher_8000_3200_8000_1
A-log-tMSE+$\Loss_0$ \cite{Wisdom2021_WhatAllFuss} & \agga & \onetwospk & \Uline{12.8} & \Uline{26.4} & \Uline{19.6} & 10.7 & 14.5 \\ %% upit_generalized_sdr/dying_zebra/ckpt_740000_SMSWsjMeetingReader_stitcher_8000_3200_8000_1
\midrule
SA-SDR&  \aggs & \onetwospk & 12.5 & 30.3 & 19.8 & 9.7 & 16.1 \\ %% upit_generalized_sdr/convinced_smelt/ckpt_830000_SMSWsjMeetingReader_stitcher_8000_3200_8000_1
SA-log-MSE&  \aggs & \onetwospk & 13.3 & \thl 31.5 & 19.3 & 11.6 & 14.7 \\ %% upit_generalized_sdr/pale_crayfish/ckpt_880000_SMSWsjMeetingReader_stitcher_8000_3200_8000_1
SA-log1p-MSE &  \aggs & \onetwospk & 15.1 & 25.1 & 18.7 & 11.4 & 15.7 \\ %% upit_generalized_sdr/testy_mockingbird/ckpt_770000_SMSWsjMeetingReader_stitcher_8000_3200_8000_1
SA-skewed-SDR &  \aggs & \onetwospk & 15.1 & 28.9 & 18.6 & 12.6 & 10.6 \\ %% upit_generalized_sdr/foolish_kite/ckpt_610000_SMSWsjMeetingReader_stitcher_8000_3200_8000_1
% SA-skewed-tSDR & \aggs & \onetwospk & 15.7 & 31.5 & 18.6 & 10.1 & 11.1 \\ %% upit_generalized_sdr/competitive_dragon/ckpt_720000_SMSWsjMeetingReader_stitcher_8000_3200_8000_1
SA-tSDR & \aggs & \onetwospk & \thl 12.2 & 30.8 & \thl 19.9 & \thl 8.2 & \thl 17.9 \\ %% upit_generalized_sdr/informal_cheetah/ckpt_670000_SMSWsjMeetingReader_stitcher_8000_3200_8000_1
SA-\tsdreps-tSDR & \aggs & \onetwospk & 12.8 & 27.5 & 19.6 & 9.1 & 16.3 \\ %% upit_generalized_sdr/distinct_locust/ckpt_880000_SMSWsjMeetingReader_stitcher_8000_3200_8000_1
    \bottomrule
\end{tabular}
    \label{tab:meetings}
\end{table}

%% Introduction
We compare the performance of models trained with different loss variants for meeting-like data in \cref{tab:meetings}.
%% Training
Training is performed on \SI{2}{\second} long segments randomly cut from the meeting-style data.
These segments can contain any number of speakers so we discard any segments with more than two speakers (that our two-output separator cannot handle) and less than one speaker.
Some losses cannot handle single-speaker segments, i.e., when one reference signal is silent.
For these losses, we additionally discard all single-speaker segments during training (roughly \SI{50}{\percent}).
The number of speakers seen during training is indicated by the \enquote{\#spk train} column in \cref{tab:meetings}.
Two-speaker segments always contain some speech in each reference signal but are not necessarily fully overlapped.
% They often contain large portions of silence and short segments of speech.

\subsubsection{Averaged SDR Losses}%%\vspace{-1em}
%% A-SDR variants
The upper half of \cref{tab:meetings} compares the different conventional averaged losses (prefixed with \enquote{A-}).
The A-log-MSE and A-SDR, as expected, show similar numbers where the A-log-MSE is slightly worse, probably due to model selection discussed in \cref{sec:log-mse}.
Modifying the log-MSE so that it can handle single-speaker training segments improves the \gls{DER} significantly while the \gls{WER} and BSSEval \gls{SDR} are unchanged.
This is expected as additional single-speaker training segments likely improve the silence estimation that is judged by \gls{DER} and \gls{saSDR} while \gls{WER} and BSSEval-SDR only judge speech regions.
A similar effect can be observed when switching from A-tSDR to A-\tsdreps-tSDR.
% \inred{Why is tSDR not better than SDR?}
% \fromcb{If I remember correctly from TCL, the SDR gets unstable after some time. Hence you can train longer with tSDR. I guess, you didn't trained until both models are fully converged.}
% The best \gls{WER} (ignoring the silent regions) among the averaged loss variants can be achieved with the A-\tsdreps-tSDR that limits the value range of the loss and is trained on single- and two-speaker segments.
A-\tsdreps-tSDR and log-tMSE+$\Loss_0$ achieve the best performance in speech regions where  A-\tsdreps-tSDR is preferable because it does not require switching the loss function depending on the energy of the reference signals.
The best overall performance among the averaged loss variants, including silence evaluation, can be achieved with A-log1p-MSE.

The scale-dependent skewed \gls{SDR} does not seem to be well-suited for training a source separation system.
We can observe that the loss pushes the outputs towards silence: The attenuation ratio and \gls{DER} are relatively good while all other metrics show a poor performance.

\subsubsection{Source-Aggregated SDR Losses}%%\vspace{-1em}
%% SA-SDR variants
Comparing the averaged losses with the source-aggregated losses, we can observe a consistent improvement for most loss variants.
% We can observe a similar behaviour within the source-aggreagted losses as we observed in the averaged losses: SA-SDR is slightly better than SA-log-MSE.
% 
%% Modifications that distort the loss are no longer beneficial
Many modifications of the standard SDR from the upper half of \cref{tab:meetings} improve the performance when averaged because they allow training on single-speaker segments.
They, however, distort the loss values, trading-off between more realistic data and an undistorted loss.
For the source-aggregated losses that always allow single-speaker segments, they lose the benefit of better training data and most of them degrade the performance (compare \enquote{log-1p} and \enquote{\tsdreps-tSDR} losses).
The SA-log1p-MSE variant, for example, is now slightly worse than the SA-log-MSE because of the constant $1$.

The overall best performance on meeting-style data can be achieved with the SA-tSDR.
It is more elegant and easier to use than the A-\tsdreps-tSDR or the A-log-tMSE+$\Loss_0$ because it has fewer hyperparameters to tune.
Very close performance can be achieved in all metrics by the SA-SDR loss that has no hyperparameters.

We observed that the averaged loss variants often become unstable late in training, probably because they focus the best separated outputs.
The source-aggregated loss variants that better balance different outputs did not become unstable in our experiments.

\section{Conclusions}

We compared different \gls{SDR}-based objective functions for source separation that allow training a neural-network-based separator on more realistic data, including silent reference signals for single outputs.
We found that stabilizing losses for perfect reconstruction and allowing silent targets, which result in training data closer to the evaluation data, often improves the performance.
We proposed a novel way of combining the \glspl{SDR} computed on individual output channels that elegantly addresses the problems of conventional \gls{SDR}-based losses and improves the performance of trained models.

\section{Acknowledgements}
Computational resources were provided by the Paderborn Center for Parallel Computing.
C. Boeddeker was supported by DFG under project no. 448568305. 

\clearpage
\balance
\bibliographystyle{IEEEbib}
\bibliography{references}

\end{document}